# Laser-guided lightning

**Author list**

Aurélien Houard[1,*], Pierre Walch[1], Thomas Produit[2,†], Victor Moreno[2], Benoit Mahieu[1], Antonio Sunjerga[3], Clemens Herkommer[4], Amirhossein Mostajabi[3], Ugo Andral[2], Yves-Bernard André[1], Magali Lozano[1], Laurent Bizet[1], Malte C. Schroeder[2], Guillaume Schimmel[2], Michel Moret[2], Mark Stanley[5], W. A. Rison[5], Oliver Maurice[6], Bruno Esmiller[6], Knut Michel[4], Walter Haas[7], Thomas Metzger[4], Marcos Rubinstein[8], Farhad Rachidi[3], Vernon Cooray[9], André Mysyrowicz[1,10], Jérôme Kasparian[2,11], Jean-Pierre Wolf[2,*]

**Affiliations**

[1]Laboratoire d'Optique Appliquée – ENSTA Paris, Ecole Polytechnique, CNRS, IP Paris - 91762 Palaiseau, France

[2]Groupe de Physique Appliquée, Université de Genève, Ch. de Pinchat 22, 1211 Geneva 4, Switzerland

[3]EMC Laboratory, Electrical Engineering Institute, Ecole Polytechnique Fédérale de Lausanne (EPFL), Lausanne, Switzerland

[4]TRUMPF Scientific Lasers GmbH + Co. KG, Feringastr. 10a, 85774 Unterföhring, Germany

[5]Langmuir Laboratory for Atmospheric Research, New Mexico Institute of Mining and Technology, Socorro, NM, USA

[6]ArianeGroup, 51 route de Verneuil, 78130, Les Mureaux, France

[7]Swisscom Broadcast AG, Ostermundigenstrasse 99, 3050 Bern, Switzerland

[8]School of Management and Engineering Vaud, University of Applied Sciences and Arts Western Switzerland, 1401 Yverdon-les-Bains, Switzerland

[9]Department of Electrical Engineering, Uppsala University, Uppsala, Sweden

[10]André Mysyrowicz Consultants, 6 Rue Gabriel, 78000 Versailles, France

[11]Institute for Environmental Sciences, Université de Genève, Bd Carl Vogt 66, 1211 Geneva 4, Switzerland

[†]Present address: Institute of Materials Research and Engineering, Agency for Science Technology and Research (A*STAR), Singapore, 138634 Singapore

[*]Corresponding authors; Email: aurelien.houard@polytechnique.edu, jean-pierre.wolf@unige.ch

**Abstract**

**Lightning discharges between charged clouds and the earth's surface are responsible for considerable damages and casualties. It is therefore important to develop better protection methods in addition to the traditional Franklin rod. Here we present the first demonstration that laser-induced filaments, formed in the sky by short and intense laser pulses can guide lightning discharges over considerable distances. We believe that this experimental breakthrough will lead to progress in lightning protection and lightning physics. An experimental campaign was carried out on the Säntis Mountain in Northeastern Switzerland during the Summer of 2021 with a high repetition rate terawatt laser. The guiding of an upward negative lightning leader over a distance of 50 m was recorded by two separate high-speed cameras. The guiding of negative lightning leaders by laser filaments was corroborated in three other instances by VHF interferometric measurements, and the number of X-ray bursts detected during guided lightning events was significantly increased. While this research field has been very active for more than 20 years, this is the first field-result that experimentally demonstrates lightning guiding by lasers. This work paves the way for new atmospheric applications of ultrashort lasers and represents a significant step forward in the development of a laser based lightning protection for airports, launchpads or large infrastructures.**

**Main text**

Lightning has fascinated and terrified humankind since time immemorial. Based on satellite data, the total lightning flash rate worldwide, including cloud-to-ground and cloud lightning, is estimated to be between 40 and 120 flashes per second[1], causing considerable damage and casualties. The documented number of lightning fatalities is well above 4,000 (ref. 2) and lightning damages amount to billions of dollars every year[3]. The most widely used external protection against direct lightning strikes is still the lightning rod, also known as Franklin rod or lightning conductor. The lightning rod, whose invention in the 18[th] century is attributed to Benjamin Franklin, consists of a pointed conducting mast connected to the ground. It protects buildings and their immediate surroundings by providing a preferential strike point for the lightning and guiding its electric current safely to the ground.

A method to initiate lightning discharges with a small rocket trailing a long, grounded conducting wire was demonstrated by Newman *et al.* in 1965[4]. In contrast to the classical lightning rod, which is intended to be struck by lightning that approaches the protected structure, the rocket-and-wire technique is intended to trigger lightning artificially. Rapidly inserting a wire into the strong electric fields near the ground below a thundercloud results in a field at the tip of the wire sufficiently enhanced to produce electrical breakdown. If the small rocket is fired at the right moment, when conditions for lightning are met, this method can initiate lightning with a success rate of up to 90%[5]. However, it requires expendable rockets and wires, the falling debris of which presents a danger.

The idea of using a laser to trigger lightning was first suggested by Ball[6]. A first attempt to trigger and guide natural lightning with lasers was made by Uchida *et al.* in 1999 using a combination of three lasers with a kilojoule energy to form a 2-meter long plasma spark[7-9]. In this article, we present results of a campaign relying on the use of laser filamentation. The principle of the filament lightning rod is the following: intense and short laser pulses are sent toward the clouds. During their propagation, they undergo a filamentation process[10-12]. The laser pulse first shrinks in size because of the laser induced change of refractive index of air, acting like a self-generated series of increasingly converging lenses. Eventually, the laser pulse becomes sufficiently intense to ionize air molecules in a high field process. The further propagation of the laser pulse is ruled by a dynamic competition between beam self-focusing and the defocusing effect due to the presence of free electrons. This competition maintains narrow channels of ionizing laser pulses over long distances. Along these filamentary regions, air molecules are rapidly heated by the absorbed laser energy and expelled radially at supersonic speed, leaving behind long-lived channels of air with reduced density[13-17]. These low-density channels of millisecond duration have higher electronic conductivity and consequently offer a privileged path for electric discharges. Meters long electric discharges triggered and guided by filaments have been demonstrated in the laboratory [18-23] and they have been shown to compete successfully with traditional lightning rods[24]. The ionized length of filamentation can reach a hundred meters when the initial pulse power of picosecond duration is in the TW ($10^{12}$ W) range[25-26]. The filamentation process can be controlled so that it starts up to a km away from the laser source[26-27]. It is therefore conceivable that filamentary channels can serve to guide and possibly even to trigger lightning discharges under appropriate weather conditions.

In our experimental campaign performed during the Summer of 2021, a Yb:YAG laser emitting pulses of picosecond duration and 500 mJ energy at a wavelength of 1030 nm and at 1 kHz repetition rate[28] was installed in the vicinity of a 124-meter tall telecommunications tower located on top of the Säntis mountain, in Northeastern Switzerland (for a description of the experimental set up, see ref [29-30], Methods and Extended Data Fig. 1 and 2). This tower, which is struck by lightning about 100 times a year, is equipped with multiple sensors to record the lightning current, electromagnetic fields at various distances, X-rays, and radiation sources from the lightning discharges (for a detailed description of the lightning instrumentation, see [31-34]). The laser pulses were directed upward, with a propagation path passing in the vicinity of the tip of the tower, which is equipped with a Franklin rod (see Fig. 1). Relying on the results of a preliminary horizontal propagation campaign in laboratory, the laser conditions were adjusted so that initiation of filamentary behavior started close to, but above the tip of the tower and had a length of at least 30 m.

Between July 21st and September 30th, 2021, the laser was operated during a total of 6.3 hours of thunderstorm activity occurring within 3 km of the tower. The tower was hit by at least 16 lightning flashes, 4 of which, denoted L1, L2, L3, and L4 (See Extended Data Table 1 for a list of the lightning events) occurred during the laser activity. All recorded lightning strikes were upward, like 97% of the strikes observed at the Säntis Tower since 2010[35]. However, remarkably, while observations at the Säntis tower over 9 years in the absence of laser show 84% negative, 11% positive and 5% bipolar flashes[36], the four recorded laser events were all positive flashes, connecting the top of the tower to a positive charge center in the cloud (according to the atmospheric electricity sign convention, a positive (negative) lightning flash is produced by a positively (negatively) charged cloud and generates a positive (negative) background electric field around the tower. It can then induce upward negative (positive) leaders). Only one of these 4 laser events (L2) occurred during a relatively clear sky on July 24 2021, at 16:24 UTC, allowing the recording of the path of the lightning discharge viewed from two directions with two high speed cameras located, respectively, 1.4 and 5 km from the tower. Snapshots of this event are displayed in Fig. 2. They show that the lightning strike initially follows the laser path over most of the initial 50 m distance (see also Extended Data Figure 3). Notice that the discharge is not completely straight along this initial segment, as it would be in the case of triggering with a rocket trailing a wire. This difference is, however, well known from laser triggered and guided discharges performed at High Voltage facilities[37]. Figure 3 shows time resolved sequences of ascending negative leaders for two upward flashes, one (L2) that occurred with the laser in operation and the other without the laser on 2.07.2019 at 00:22 UTC. Note the absence of branching during the laser guiding stage (lowest vertical section) of L2.

A lightning storm is the source of emission of electromagnetic waves spanning a broad frequency range from radio waves to gamma-rays. VHF activity (of 1-10 meters wavelength) is particularly useful for the study of discharges during their formation stage. A VHF interferometer system developed by New Mexico Tech was installed during the 2021 Summer measurement campaign in the vicinity of the Säntis tower[38]. This system consists of inverted V-shape antennas recording the phase differences between incoming VHF radiation sources due to their different locations. By using a cross correlation algorithm[39], the location of the source can be obtained in a 2D space (azimuth, elevation). The system was capable of tracking the lightning leader propagation with a spatial resolution of several meters and a time resolution on the order of microseconds. Figure 4 compares the VHF sources located by the interferometer for two flashes, one (L1) with the laser active and the other without the laser (N07). In the former case, an accumulation of radiation sources located along the laser path is observed over a distance of approximately 60 m (other events are presented in Extended Data Fig. 4). As presented in Extended data Table 2 and Figure 5, the standard deviation of the distance from the sources to the laser is reduced by 45% over the same 60 m when the laser is on.

The lightning current, electric fields and X-rays were observed for the L1, L3 and L4 events. Results for L1 and L3 are presented in Fig. 5 and compared to an event with the same positive polarity measured in the absence of laser. The X-ray detector field of view was pointing toward the laser trajectory. The three events presented here were compared to two other positive upward events during 2021 prior to the laser installation. They exhibited similar current and electric field waveforms in terms of amplitudes and step intervals. On the other hand, the number of X-ray bursts in the presence of the laser beam (4.3 per event) was significantly higher than in its absence (1 per event). Note that most of the X-ray bursts observed with the laser on are detected during the time corresponding to the laser-guided leader propagation (first 500 μs). Unpublished experiments performed at the Laboratoire d'Optique Appliquee, show that meter-long guided discharges emit an X-ray burst in the forward direction. This suggests that these X-ray bursts were emitted during straight sections of the discharge (see Fig. 2).

**Discussion of the results**

Before the 2021 campaign described in this paper, a few attempts at guiding and/or initiating lightning using short laser pulse filaments with TW peak power were made in New Mexico in 2004[40], and in Singapore in 2011. These earlier campaigns failed to produce evidence of laser guiding or initiation of lightning discharges. This raises the following two questions: (1) why was the Säntis campaign more successful than the two prior attempts? and (2) why did only upward negative leaders (associated with upward positive flashes) exhibit guiding from the laser during the 2021 Säntis campaign?

We conjecture that an important factor contributing to the success of the Säntis campaign is the repetition rate of the laser, which was higher by two orders of magnitude when compared to previous attempts. Prior to a lightning flash at Säntis, the electric field is typically varying very slowly (tens to hundreds of milliseconds). This is because most of the flashes are self-initiated[41]. Using a kHz repetition rate allows therefore intercepting all the lightning precursors developing above the tower. In addition, during filamentation, a small fraction of the free electrons created by high field ionization is captured by neutral oxygen molecules. At high laser repetition rates, these long-lived charged oxygen molecules accumulate, keeping a memory of the laser path[42]. Electrons captured by neutral oxygen molecules have a trapping potential of 0.15 eV[43] instead of 13.62 and 15.58 eV for electrons bound to oxygen and nitrogen molecules, respectively, and they can therefore easily be set free by heat or inelastic collisions with energetic electrons accelerated by the ambient electric field. Laboratory experiments have provided evidence of the presence of such free electrons in filaments at kHz laser repetition rate[44]. Filaments with an accumulation of positively and negatively charged molecules as well as free electrons form a polarizable medium. Charge migration inside the filament strings, induced by the atmospheric electric field, can produce a reinforcement of the electric field, promoting the formation of discharge segments[45]. Additional campaigns and more theoretical work are necessary to confirm this conjecture.

In order to explain why only negatively charged ascending leaders (positive lightning flashes) were observed in the presence of the laser beam, it is instructive to consider the electric field conditions required to generate a discharge bridging the gap between the lower tip of the filamentary path and the metallic rod at the top of the tower (see details on the model in the Methods). Consider first the case where the background electric field is generated by positive charges in the cloud. In this case the gap could be bridged either by negative streamers emanating from the tower tip or by positive streamers generated by the lower tip of the laser filament. Our calculations, based on the conditions necessary for the initiation and propagation of streamer discharges and detailed in the supplementary material, show that bridging the gap by positive streamers emanating from the bottom tip of the filament takes place at a lower background electric field (that we denote $E_{pos\text{-}filament}$) in comparison to the electric field needed to bridge the gap by negative streamers. Note that this scenario will give rise to a positive lightning flash.

Now, let us consider the case where the background electric field is generated by negative charges in the cloud. Again, two scenarios are possible. The gap could be bridged either by positive streamers initiated from the tower tip or by negative streamers initiated from the lower tip of the filament. As before, positive streamers, emanating now from the tower, bridge the gap at a lower electric field denoted $E_{pos\text{-}tower}$, giving rise to negative lightning flashes. Furthermore, our calculations show that $E_{pos\text{-}filament} < E_{pos\text{-}tower}$. This means that positive lightning flashes are more likely to be initiated in the presence of the laser filaments compared to negative flashes. Note that the same tendency has been observed in the laboratory, where the triggering of meter scale discharges by laser filaments was studied with positive and negative polarities[24].

In conclusion, the results of the Säntis experimental campaign in the summer of 2021 provide circumstantial evidence that filaments formed by short and intense laser pulses can guide lightning discharges over considerable distances. These preliminary results should be confirmed by additional campaigns with new configurations. The use of a Franklin rod with a minimum distance from the laser path could increase the probability to guide lightning flashes of both polarities. The use of visible laser wavelength (obtained by second harmonic generation) could also increase the guiding efficiency of the filament. Additionally, note that, based on the results presented in this paper, displacing the onset of filamentation toward charge centers in the cloud might result in an increase in the guiding capability of the laser or even the initiation of lightning discharges. This is the subject of future experimental work.

**Acknowledgments**

We acknowledge the strong support from M. Pittman, Y. Bertho, IJCLab and University Paris Saclay, who allowed us to perform preparatory experiments with the laser. We also acknowledge the Swiss federal office of Civil Aviation and Swisscom Broadcast AG for their help in the preparation of the lightning campaign at the Säntis, as well as Skyguide and Säntis-Schwebebahn AG for their active cooperation during the campaign. We acknowledge the help of Robert Bessing and Sandro Klingebiel from Trumpf Scientific Lasers for the laser installation and Christophe Vassaux, Roland Pellet and Laurent Leuenberger from the mechanical workshops of the University Geneva for the construction and installation of the telescope housing.

**Funding**

European Union Horizon 2020 Research and innovation programme FET-OPEN grant (737033-LLR): all authors

Swiss National Science Foundation (grant Nr. 200020-175594): AS, AMo, MR, FR

Swiss National Science Foundation (grant No. 200021-178926): UA, MS, JPW

French Direction Générale de l'Armement: LB, PW, YBA

B. John F. and Svea Andersson donation at Uppsala University: VC

**Author Contributions Statement**

JPW, AH, AMY, JK, FR, MR initiated and defined the project. AH, JPW, AMY, TM, BE, FR, MR obtained the funding and supervised the project. AH coordinated the project. TP, UA, GS, MM, JK, JPW, BM, PW, AH, WH designed and installed the experiment. AS, FR and MR designed and installed the lightning experiment. PW, TP, BM, VM, UA, JK, JPW, YBA, ML, LB, AH, MCS conducted the laser experiments on site. AS, AMO, MR, FR, OM, MS, WAR conducted the lightning experiment and characterization. CH, KM, TM developed the laser system. VC developed the lightning simulations. PW, AS, JPW, JK, VM, WAR analyzed the data. AMY, AH prepared the original draft and all authors critically reviewed and approved the manuscript.

**Competing Interests Statement**

Authors declare that they have no competing interests.

**Figure Legends/Captions (for main text figures)**

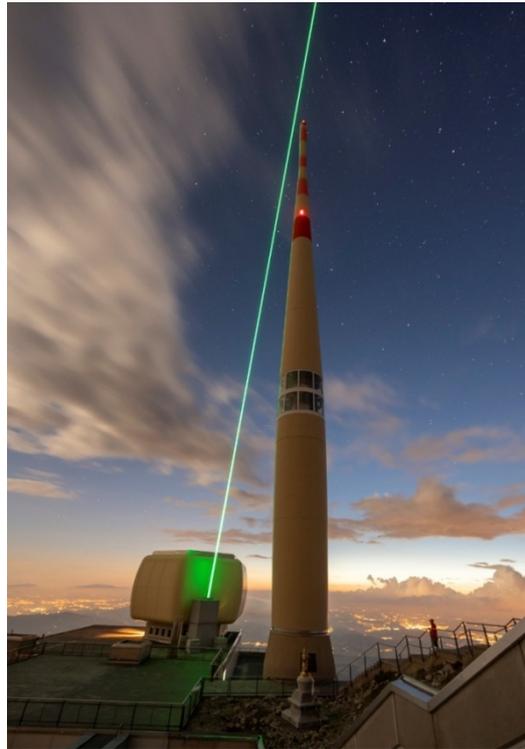

**Fig. 1: Image of the 124 m high telecommunication tower of Säntis (Switzerland).**
Also shown is the track of the laser recorded with the second harmonic of the laser at 515 nm.

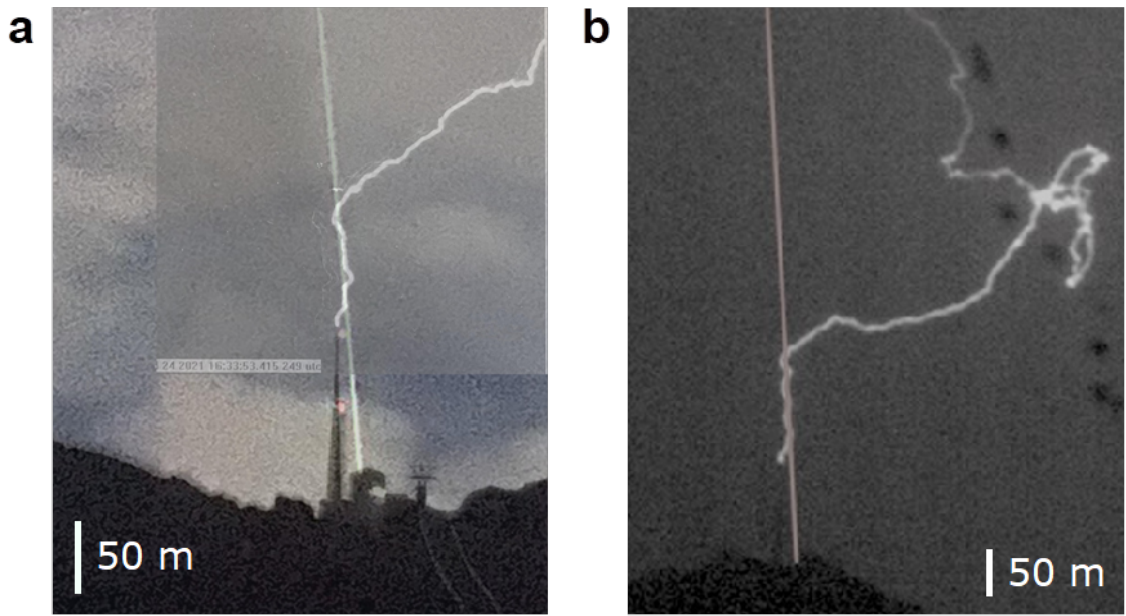

**Fig. 2: Snapshots of the lightning event of July 24 (L2) recorded in the presence of the laser.**
Snapshot recorded by the two high-speed cameras located at Schwaegalp **(a)** and Kronberg **(b).** The trajectory of the laser path taken subsequently in clear sky through second harmonic generation is also overlaid.

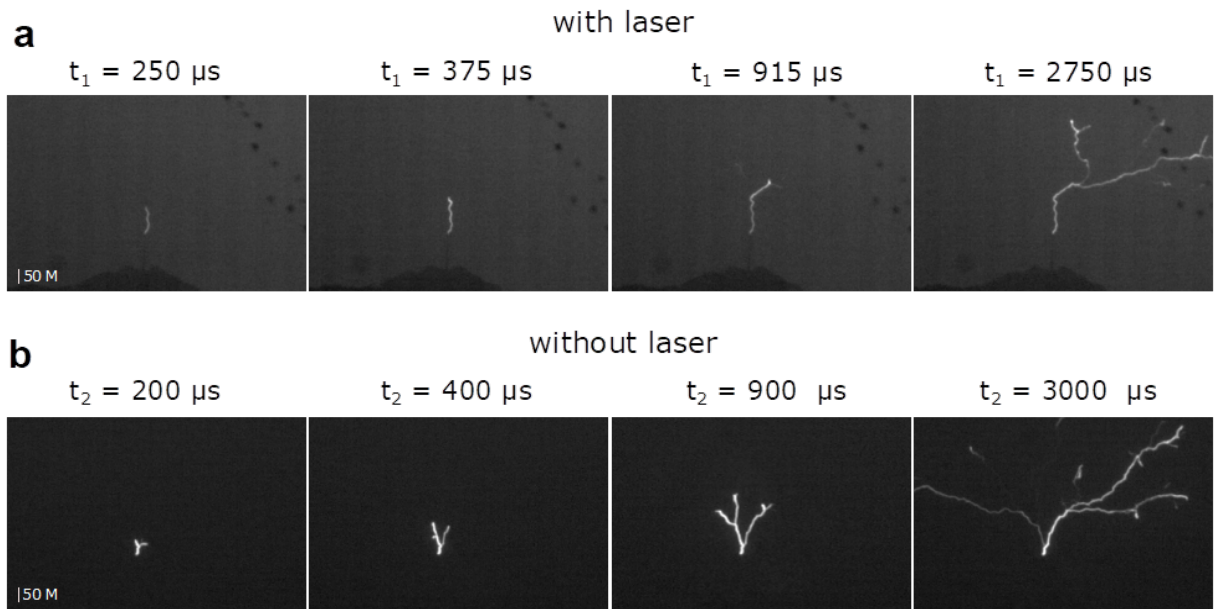

**Fig. 3: High speed camera images of upward leaders.**
**a**, Images of the lightning path in the presence of the laser recorded on July 24 2021 at (L2) 250, 375, 915, and 2750 µs after initiation of the discharge. **b**, Images of the lightning path recorded on July 2. 2019 in the absence of laser 200, 400, 900, and 3000 µs after initiation of the discharge.

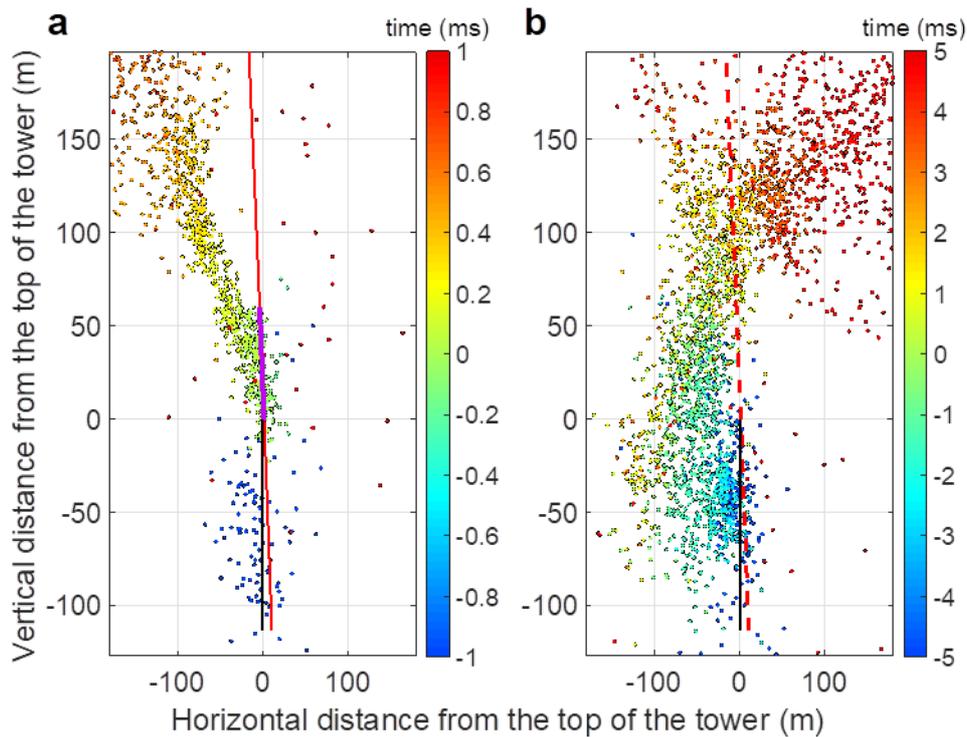

**Fig. 4: Measurements with VHF interferometer.**
2D maps of the VHF sources emitted during the lightning event L1 with the laser on (**a**) and N6 without laser (**b**). Also shown is the telecommunication tower in black and the laser path in red. Each point corresponds to a VHF emission. The color code displayed on the right corresponds to the timescale. The violet section shows the region where laser filamentation is expected.

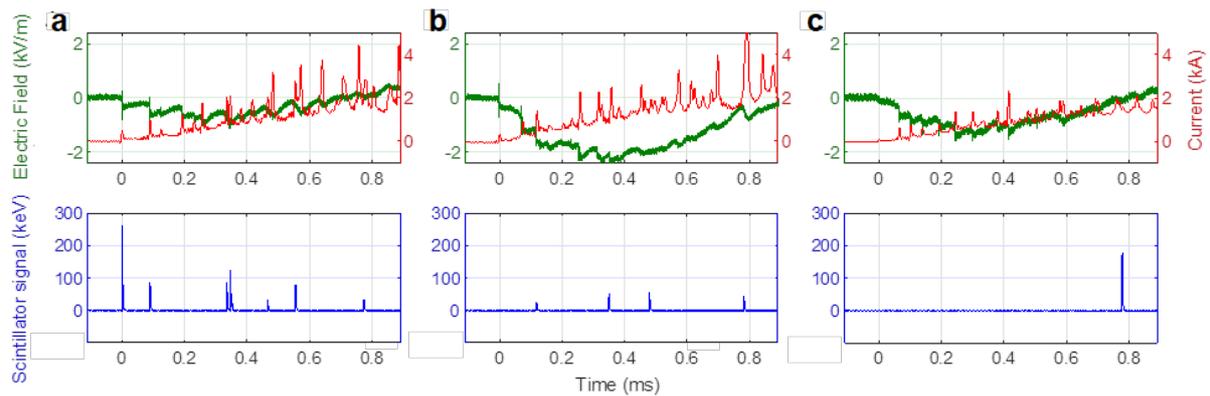

**Fig. 5: Electric signals measured for the 3 positive upward flashes L1 (a), L3 (b) and N6 (c).** Top subplot for each event: The electric field scale is given on the left y-axis and the current on the right y-axis. Bottom subplot: X-ray signal detected by the Scintillator, where each peak corresponds to the integrated X-rays energy collected during the 50 ns sampling. Events L1 and L3 correspond to events with laser and N6 to an event without laser.

# Methods

**Laser system**

The laser system operating during the campaign, fully described in [28], is a Yb:YAG Chirped pulse amplification (CPA) laser system developed by TRUMPF Scientific Lasers. It is capable of delivering laser pulses at 1030 nm with 720 mJ of energy per pulse, a pulse duration of 920 fs at a repetition rate of 1 kilohertz. Because of the configuration of our setup with a long propagation before the sending telescope, the output energy was reduced to 500 mJ and the pulse was chirped to a pulse duration of 7 ps to prevent damage on the optics. In Extended data Fig. 1b, an LBO crystal[46] was used to produce a beam at the second harmonic at 515 nm wavelength.

**Experimental setup**

The campaign was carried out at the top of the Säntis Mountain (2502 m altitude), in Northwestern Switzerland, on the summit of which stands a 124 m tall telecommunication tower. A general view of the experimental setup is presented in Extended Data Fig. 1. The laser system was located in the radome building, sheltered in an air-tight, air-conditioned and thermally isolated tent. After exiting the tent, the laser output was directed downwards by a conduit through the radome wall to the terrace, where a 4" folding mirror directed the beam into a beam-expanding sending telescope featuring a 7.14 magnification ratio. The entire laser path toward the telescope, presented in Extended data Fig. 1a, was protected by an isolated aluminum housing to prevent any beam leakage and to reduce the perturbation from the environment.

The telescope was composed of an additional folding mirror, a secondary 100 mm spherical mirror and a 430-mm diameter off-axis aspheric (elliptic) primary mirror[29]. The beam output, which had a diameter of 250 mm, was sent toward the tower tip with a vertical angle of 7°. Translation stages on the secondary mirror allowed us to focus the beam near the tower tip in order to set the onset of the filamentation process in the desired area in which upward lightning is initiated. The focal length of the telescope was set to 150 m in order to produce a dense filamentation area of 30-50 m above the tower tip.

Note that during all laser operation time, the airspace was closed by the air-traffic authority. Furthermore, air traffic was monitored by an ADS-B transceiver automatically switching off the laser in case of aircraft incursion into the temporary closed airspace zone.

**High speed camera measurements**
On July 24, the L2 event was observed with the two high speed cameras, one operating at 24 000 frames per second, installed on the Kronberg mountain, and the second operating at 10 000 frames per second installed at Säntis Das Hotel (Schwaegalp). Note that high-speed camera records of upward positive flashes are very rare and there are only a few reported in the literature[47-48].

Figure 2 displays two representative frames taken from the two fast cameras. In order to precisely calibrate the position of the laser, several comparative procedures were carried out: images from the fast cameras in daylight to identify the position of the tower and the surrounding topography, high resolution pictures at night with a D810 Nikon next to the fast cameras when the laser was operating, and reconstruction of the laser direction and position using precise GPS data. In the two pictures in Fig. 2, depicting an upward positive flash, an initial segment of about 70 m from Schwaegalp and 120 m from Kronberg is observed, following the path of the laser beam.

Notice that the discharge is not completely straight along this initial segment, as it would be in the case of triggering with a rocket trailing a wire. This difference is, however, well known from laser triggered and guided discharges performed at High Voltage facilities[35]. The reason is that the current displacement is much more complex in a distribution of moving charges than along a wire. For instance, the moving charges create space charges that locally screen the electric field.

For events without laser, individual images from one of the cameras were available, allowing to plot histograms of distances to the laser beam as projected in the plane perpendicular to the camera line of sight (Extended data Fig. 3). The difference in behavior between the event with (L2) and without laser (N05, N08) is apparent.

**Leader velocity**
The velocity estimated from the interferometer data (between $1 \times 10^5$ and $6 \times 10^5$ m/s) is consistent with estimates deduced from the 24'000 fps image sequence of the fast camera located in Kronberg, that yields a velocity decaying from $4 \times 10^5$ m/s when leaving the tower tip to $9 \times 10^4$ m/s at the first branching 120 m above it, with an average velocity of $2 \times 10^5$ m/s over this interval. Unfortunately, no fast camera image was available under comparable conditions (upward positive flash) without the laser during the campaign. This value is, however, comparable to the typical reported values for upward negative leaders (corresponding to positive flashes): $2 \times 10^5$ m/s for rocket-triggered lightning[49] and $1 \times 10^5$ m/s for virgin air[50]. Note that from the fast camera images, the propagation velocity of the branches of the laser-free event of 2019-07-02 00:22:46 is estimated to be $2 \times 10^5$ m/s, without any discernible acceleration or deceleration trend over the first 900 m.

**Modeling and simulation of the effect of the filamentation on the lightning flashes initiation**
In order to understand the conditions necessary for the initiation of a lightning flash in the presence of the laser filament, we will describe in what follows the physical processes that are involved[51,52]. We will start by presenting the standard nomenclature used in the lightning literature. When the electric field in air exceeds the breakdown electric field, the free electrons in the air start ionizing other atoms and molecules, giving rise to what is commonly known as an electron avalanche. As the avalanche continues to grow, a stage will be reached where the accumulated space charge becomes so large that it starts creating electric fields comparable to, or higher than the background electric field. At this stage the electron avalanche becomes a self-propagating electrical discharge known as a streamer. The streamers can propagate in background electric fields lower than the breakdown values. The currents in these streamer discharges are in the range of µA to mA. Moreover, streamers are cold discharges. That is, the temperature of the gas in the discharge remains close to the ambient temperature. During the initiation of a discharge, many streamers may originate from a common root or stem. As the current from all the streamers passes through this stem, the temperature of the stem increases. As the stem heats up, thermal ionization sets in, increasing the electron density in the stem. When this electron density reaches a critical density, a rapid transfer of energy from the electrons to the neutral atoms takes place, raising the conductivity and the temperature of the gas to several thousands of degrees. This conducting channel section is called a leader. Being conducting, this channel gets polarized in the background electric field, thus enhancing the electric field at its tip. From this tip, an electron avalanche starts again, leading to the creation of a new section of the leader channel. In this way, with the aid of avalanches and streamers, the leader propagates. When a downward lightning leader reaches the ground or attaches to an upward connecting leader, a ground potential wave called return stroke propagates along the leader channel neutralizing its charge. This return stroke constitutes the visible and audible phase of the lightning strike. The processes that were just described are the main elements that are being used to describe the mechanism of lightning flashes.

Now, let us consider the problem at hand. We shall estimate whether there will be electrical breakdown between the lower end of the laser filament and the tower tip, a step which is necessary for the initiation of a lightning discharge by the laser filament. This evaluation contains many steps and it is not possible to describe all of them in detail here. The interested reader may find a full discussion in references [51-53]. However, we will describe the essential elements here while at the same time providing the references for the interested reader to go deeper into the analysis. First, for a given background electric field, the electric field in the gap between the tower tip and the lower tip of the filament is calculated using the charge simulation method. This is a standard technique to calculate the electric field in a region where the potential boundaries are given[54]. Second, the growth of electron avalanches in the high electric field region where the electric field exceeds the breakdown electric field is studied[55]. The growth of avalanches is investigated using the Townsend ionization coefficient corresponding to the ambient pressure and temperature. Third, if the number of positive ions at the electron avalanche head exceeds the critical value $10^8$, within a roughly spherical region of radius 50 µm, the avalanche is assumed to be converted to a streamer discharge[56]. Fourth, once a streamer is initiated, its propagation distance into the gap is estimated using the value of the critical electric field necessary for streamer propagation. For the propagation of positive streamers at mean sea-level pressure, a background electric field of about 500 kV/m is required and, for negative streamers, the required field is higher, 1 MV/m to 2 MV/m[57-58]. Both these values have to be scaled by a factor of 0.75 to take into account the 0.75 atm atmospheric pressure at the 2500 m altitude of the Säntis tower. Once the extension of the streamers in the gap is estimated, the charge associated with the streamer burst is estimated following the procedure used by Becerra and Cooray[59]. If the charge in the streamer burst is larger than 1 µC, the streamer burst will give rise to a leader discharge[56]. Leaders can propagate in fields larger than about 150 – 200 kV/m which is much lower than the threshold necessary for streamer propagation[57,58,60]. Based on this procedure, the analysis to be presented below is carried out.

Consider the location of a laser filament above the tip of the tower. The geometry relevant to the calculation is shown in Extended Data Fig. 6. In the case of a laser filament fully polarized by the background electric field, the conditions necessary for the initiation of a lightning flash are the following. First, a streamer burst has to be initiated from the lower tip (the tip closest to the tower) of the laser filament. This streamer burst will propagate towards the tower tip and, if the average electric field over the streamer length is below the threshold field necessary for streamer propagation, it will stop before bridging the gap. As mentioned earlier, this threshold is different for negative and positive streamers. If the average electric field between the lower tip of the laser filament and the tower tip is larger than the threshold field necessary for streamer propagation, the streamer burst will travel across the whole gap distance between the laser tip and the tip of the tower. This situation is called the final jump condition. In this case, the electrical breakdown between the laser filament and the tower tip is unavoidable, and this generates the conditions necessary for the initiation of a lightning flash. If the average electric field in the gap is below the streamer propagation threshold, still a leader could be generated at the tip of the laser filament and this leader could bridge the gap between the filament and the tower. It is important to point out that if positive streamers from the laser filament cross the gap and initiate a lightning flash, the result will be a positive lightning flash. If negative streamers from the laser filament cross the gap and initiate a lightning flash, the resulting lightning flash will bring negative charge to ground, i.e., a negative lightning flash.

Based on the criteria outlined in the previous paragraph, the magnitude of the background electric fields necessary to initiate (a) a laser-assisted positive lightning flash, (b) a laser-assisted negative lightning flash, and (c) a tower-initiated negative lightning flash without the assistance of the laser filament are estimated. The results are shown as a function of the gap length between the laser filament and the tower tip and for four different lengths of the laser filament in Extended Data Fig. 7. Note that the polarity of the background electric field necessary to initiate a positive lightning flash is opposite to the background electric field necessary to initiate a negative lightning flash. First, observe that the background electric field necessary for the generation of a laser-assisted positive lightning flash is always below the background electric field necessary to initiate a laser assisted negative lightning flash. Second, observe that at background electric fields that are large enough to initiate a laser-assisted negative lightning flash, the tower itself is capable of generating negative lightning flashes without the assistance of the laser filament. This also shows that the presence of the laser filament will not significantly change the number of negative flashes striking the tower, but it will make the initiation of positive lightning flashes possible at electric field values which would not allow them in the absence of the laser filament. We have also analyzed the maximum gap distance where the laser beam is capable of influencing the lightning initiation. For laser filament lengths equal to 10, 20, 30, 40 and 50 m, the maximum gap lengths where the laser can influence the lightning initiation are 6, 7.5, 20, 22 and 25 meters, respectively. The reason why the gap length increases significantly for laser filament lengths larger than 30 m is because for these laser filament lengths, positive leaders are initiated at the laser filament tip and the breakdown is mediated by positive leaders.

**Data availability**

The datasets generated during and/or analysed during the current study are available from the corresponding author on reasonable request.

**Methods-only references**

# Extended data figures and tables

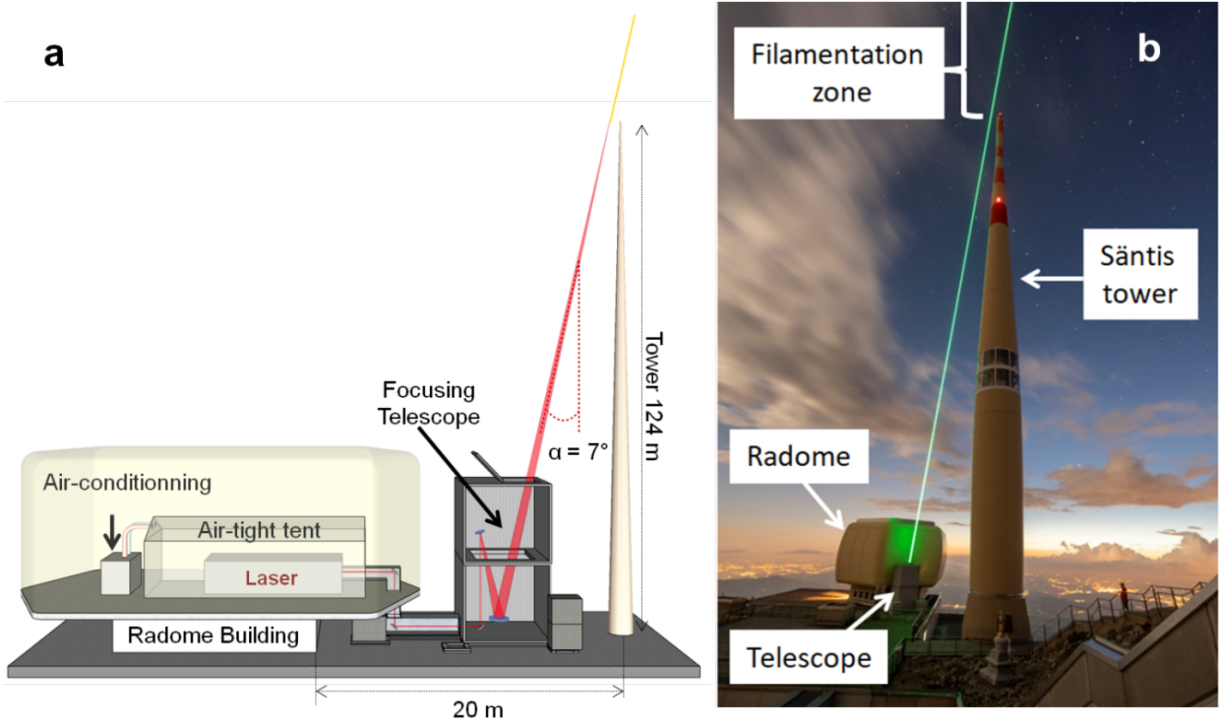

**Extended data Fig. 1 Experimental setup.**
**a**, Layout of the experimental setup on top of the Säntis Mountain. **b**, Photography of the experiment with the second harmonic of the laser beam used to visualize the laser path.

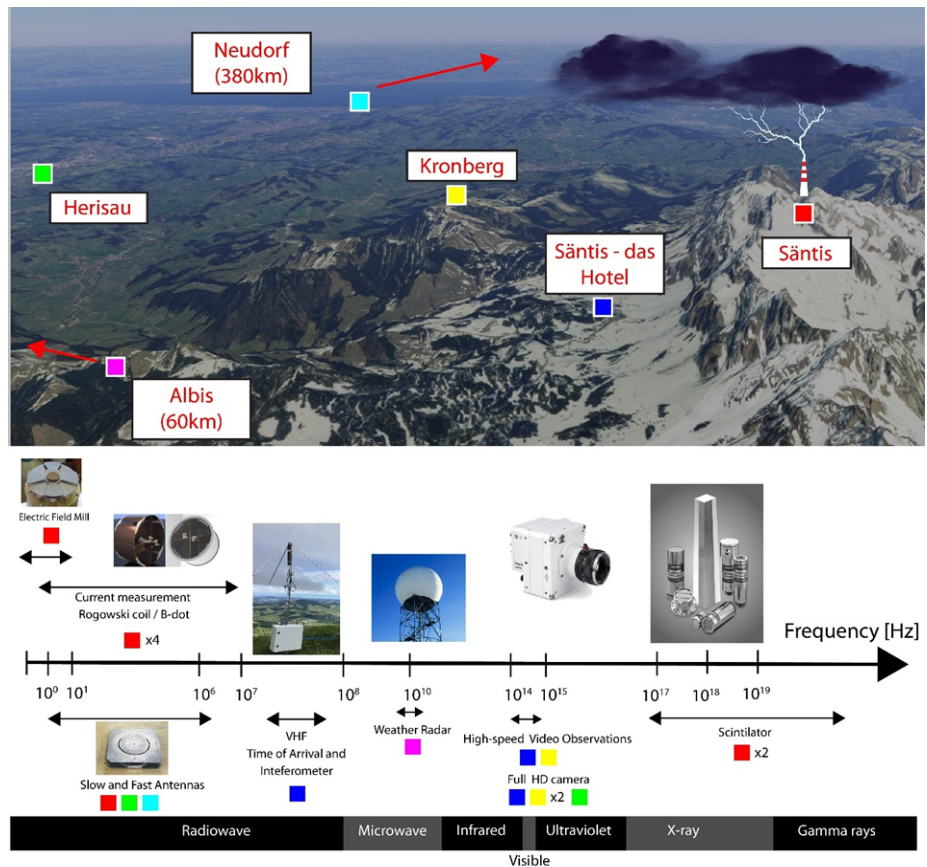

**Extended data Fig. 2 Locations of the different measuring equipments.**
The Säntis lightning measurement system consists of two fast electric field sensors, a field mill, three full HD cameras, two high speed cameras and two X-rays sensors. Data from the weather radar covering the Säntis Tower area are also made available by MeteoSwiss. Lightning currents of strikes to the tower were measured by a set of Rogowski coils and B-dot sensors located at two different heights along the tower (see[32,34] for more information on the Säntis tower instrumentation). The lightning activity was detected, located in azimuth and elevation, and GPS-time-stamped by a radiofrequency interferometer in the 1 – 160 MHz band[38] whose upper cutoff was limited to 84 MHz to avoid interference from FM radio transmitters in the area. We checked that, without electric activity in the atmosphere, the interferometer is fully insensitive to the laser system and to the laser induced filaments. An X-ray detector located in the radome measured the X-rays emitted in the range 20 keV-1 MeV. Electric field measurements were performed 20 m and 15 km away from the tower. Finally, two high-speed cameras were installed from two viewing angles to provide direct imaging of the lightning strikes in case of clear weather under elevated thunderclouds. Located in Schwägalp and Kronberg, they operated, respectively, at 10,000 and 24,000 frames per second.

**Extended data Table 1. Lightning strike events during the campaign, and available diagnostics for each event.**
During the experimental campaign, from 21 July to 30 September, 16 flashes were recorded by the current measuring system at the Säntis tower. Due to both the harsh experimental conditions as well as intrinsic instrument limitations (optical visibility, buffer size, and electromagnetic perturbations) each event could not be recorded by all instruments. The status of the laser system and each measuring system, including HSC (high-speed camera), INT (Interferometer), X-ray (X-rays sensors), E-field (20 m) (electric field 20 m from the tower) and E-field (electric field 15 km from the tower), for each of these events is given in Table 1.

Out of the 16 flashes, 5 were positive flashes (POS) and 11 were negatives flashes (NEG). Only 3 were recorded by at least one high-speed camera (two without laser and one with the laser on), while all but one were recorded by the interferometer. All investigated lightning strikes were of the upward type, like 97% of the strikes observed at Säntis since 2010[36]. All four recorded laser events were positive; by contrast, 85% negative flashes, 12% positive flashes, and 3% bipolar flashes were recorded during the Summer of 2021. Assuming conditions during the campaign similar to the average of the 10 previous years, the excess of positive flashes when the laser is on is significant beyond $10^{-3}$ (p-value=$6.7 \times 10^{-6}$ obtained with unilateral chi-2 test).

| Events | Date | UTC Time | Laser | Current | Flash polarity | HSC | INT | X-RAY | E (20m) | E (15km) |
|---|---|---|---|---|---|---|---|---|---|---|
| L1 | 24.7.2021 | 16:06:07 | ON | YES | POS | NO | YES | YES | YES | YES |
| L2 | 24.7.2021 | 16:24:03 | ON | YES | POS | YES | NO | YES | YES | YES |
| L3 | 30.7.2021 | 18:00:10 | ON | YES | POS | NO | YES | YES | YES | YES |
| L4 | 30.7.2021 | 18:02:40 | ON | YES | POS | NO | YES | NO | NO | YES |
| N01 | 30.7.2021 | 15:17:09 | OFF | YES | NEG | NO | YES | YES | YES | YES |
| N02 | 30.7.2021 | 15:22:29 | OFF | YES | NEG | NO | YES | NO | NO | NO |
| N03 | 30.7.2021 | 15:30:51 | OFF | YES | NEG | NO | YES | YES | YES | NO |
| N04 | 30.7.2021 | 15:35:41 | OFF | YES | NEG | NO | YES | YES | YES | YES |
| N05 | 30.7.2021 | 15:38:10 | OFF | YES | NEG | YES | YES | YES | YES | YES |
| N06 | 30.7.2021 | 18:04:53 | OFF | YES | POS | NO | YES | YES | YES | YES |
| N07 | 16.8.2021 | 02:00:27 | OFF | YES | NEG | NO | YES | NO | NO | NO |
| N08 | 16.8.2021 | 02:34:10 | OFF | YES | NEG | YES | YES | YES | YES | YES |
| N09 | 16.8.2021 | 05:53:24 | OFF | YES | NEG | NO | YES | YES | YES | YES |
| N10 | 16.8.2021 | 10:16:32 | OFF | YES | NEG | NO | YES | NO | NO | NO |
| N11 | 16.8.2021 | 15:06:45 | OFF | YES | NEG | NO | YES | YES | YES | NO |
| N12 | 16.8.2021 | 15:08:34 | OFF | YES | NEG | NO | YES | YES | YES | NO |

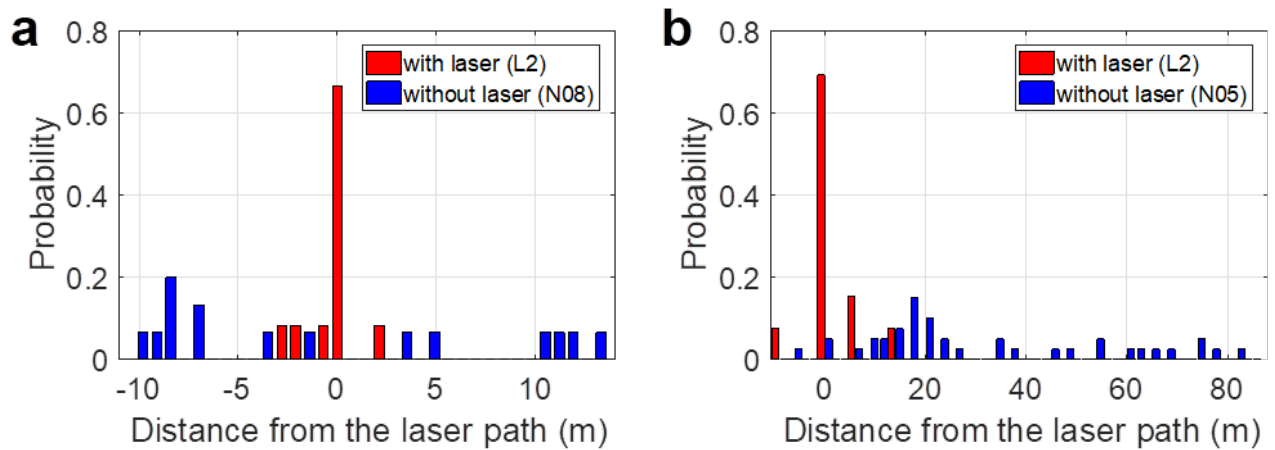

**Extended data Fig. 3 Analysis of high-speed camera measurements.**
Projected 2D histograms based on single image, comparing **a,** event L2 with N08 viewed from Säntis, and **b,** L2 with N05 viewed from Kronberg.

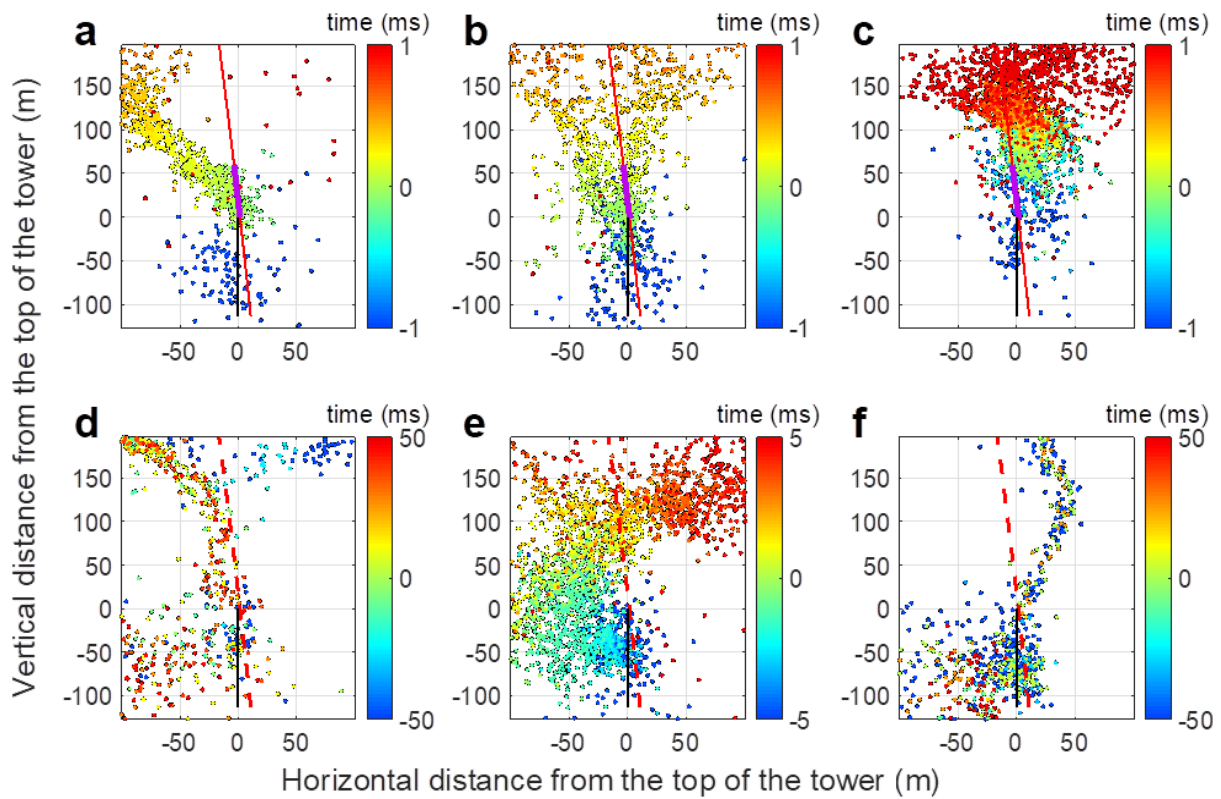

**Extended data Fig. 4. VHF sources position.**
Still images could be obtained in only one lightning strike while the laser was flashing, due to the exceptional visibility conditions required. In contrast, the RF interferometer also displays guiding for the 3 recorded events with laser (L1, L3, L4), in the form of a high density of VHF sources close to the laser beam. Typical results with laser (events L1, L3, L4) are presented respectively in panels **a-c.** Events without laser (N02, N06, N07) in panels **d-f**. Each panel displays 2D (vertical and horizontal distances from the top of the tower) maps of the sources constituting lightning strikes on the tower. The color code displayed on the right corresponds to the timescale. The Säntis tower is represented on each plot in black, the laser path in red, and the part of the laser path over which filaments are expected in violet. The dashed red line in the lower panels represents the path that the absent laser beam would have followed.

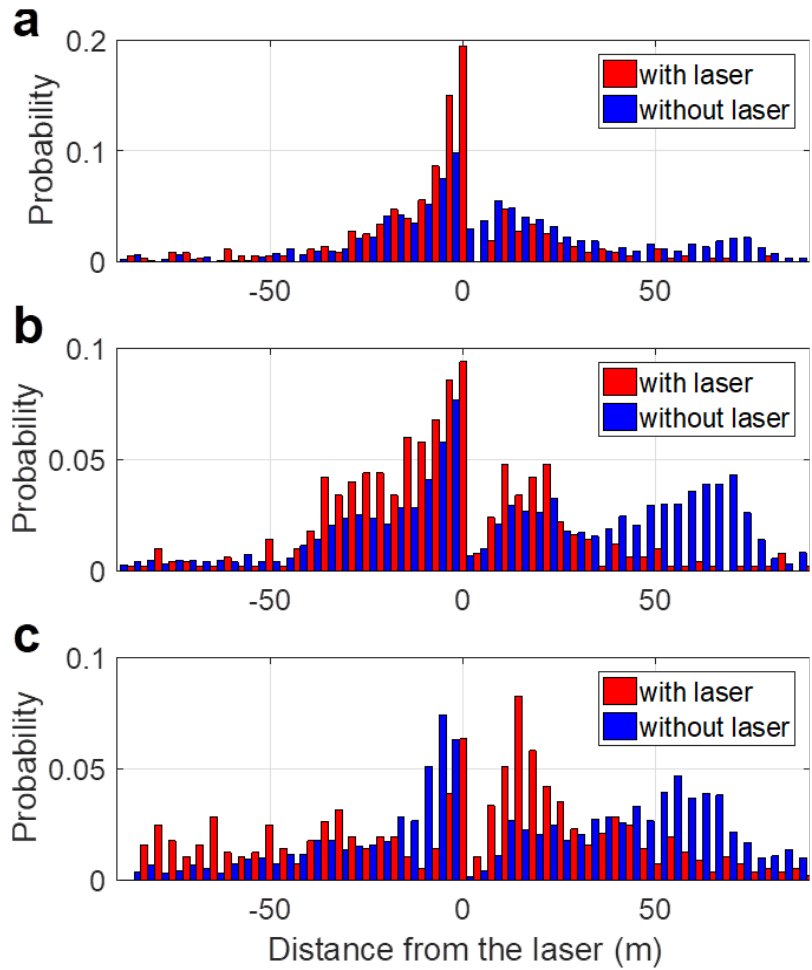

**Extended data Fig. 5**. **Histograms of VHF source distance to the laser beam.**
Comparison of the cumulated histograms of distance to the laser beam per height slice of 30 m for all events with and without laser. **a**, **b** and **c** correspond respectively, to the height slices 0-30 m, 30-60 m, and 60-90 m. When the laser is on, the distribution is centered closer to the laser beam, and narrower, as compared to when the laser is off.

**Extended data Table 2. Standard deviation of the VHF source distance to the laser beam**. Using the data presented in Extended data Fig. 5, we calculated the standard deviation of this distance for different elevations above the tower. For all 3 height slices, we performed a unilateral F-test. For the whole 0-90 m range, the standard deviations of the histograms are significantly different (P-value $\ll 10^{-15}$), that is, histograms with laser off are more dispersed. Considering the 0-60 m range, we find that the standard deviation of the distance of the sources to the laser is reduced by 45% when the laser is on. Above 60 m, this reduction due to the laser filament is much weaker, as expected since the length of the filament is estimated to be 50-60 m.

| Elevation (m) | laser | Number of sources | Standard deviation (m) |
|---|---|---|---|
| 00-30 | OFF | 1644 | 39.7 |
|  | ON | 586 | 24.2 |
| 30-60 | OFF | 1502 | 46.0 |
|  | ON | 778 | 26.4 |
| 60-90 | OFF | 1674 | 44.1 |
|  | ON | 738 | 39.6 |
| 00-60 | OFF | 3146 | 42.7 |
|  | ON | 1364 | 25.5 |

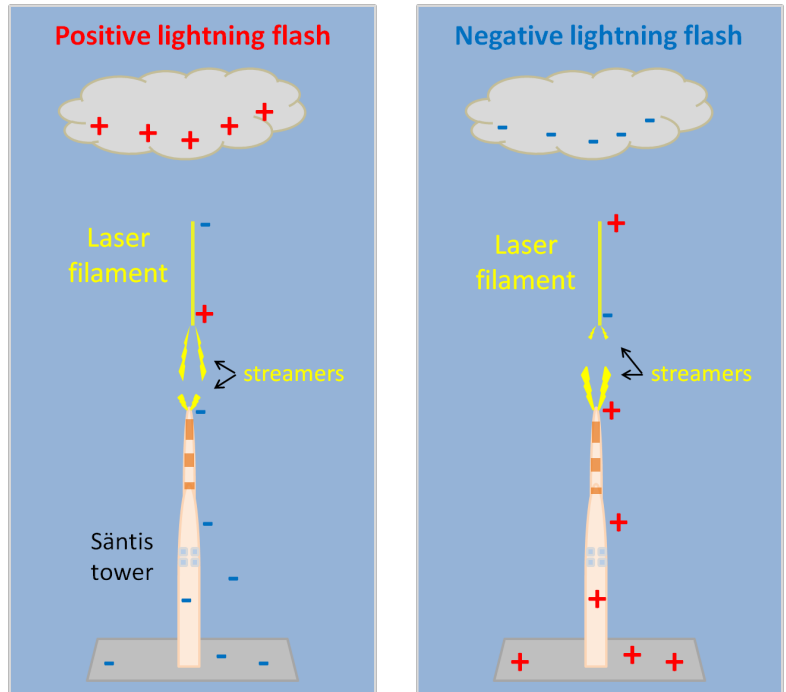

**Extended data Fig. 6. Geometry considered for the simulation.**
The laser filament is assumed to be located vertically and directly above the tower tip. Left panel: conditions associated with a positive flash. Right panel: conditions associated with a negative flash.

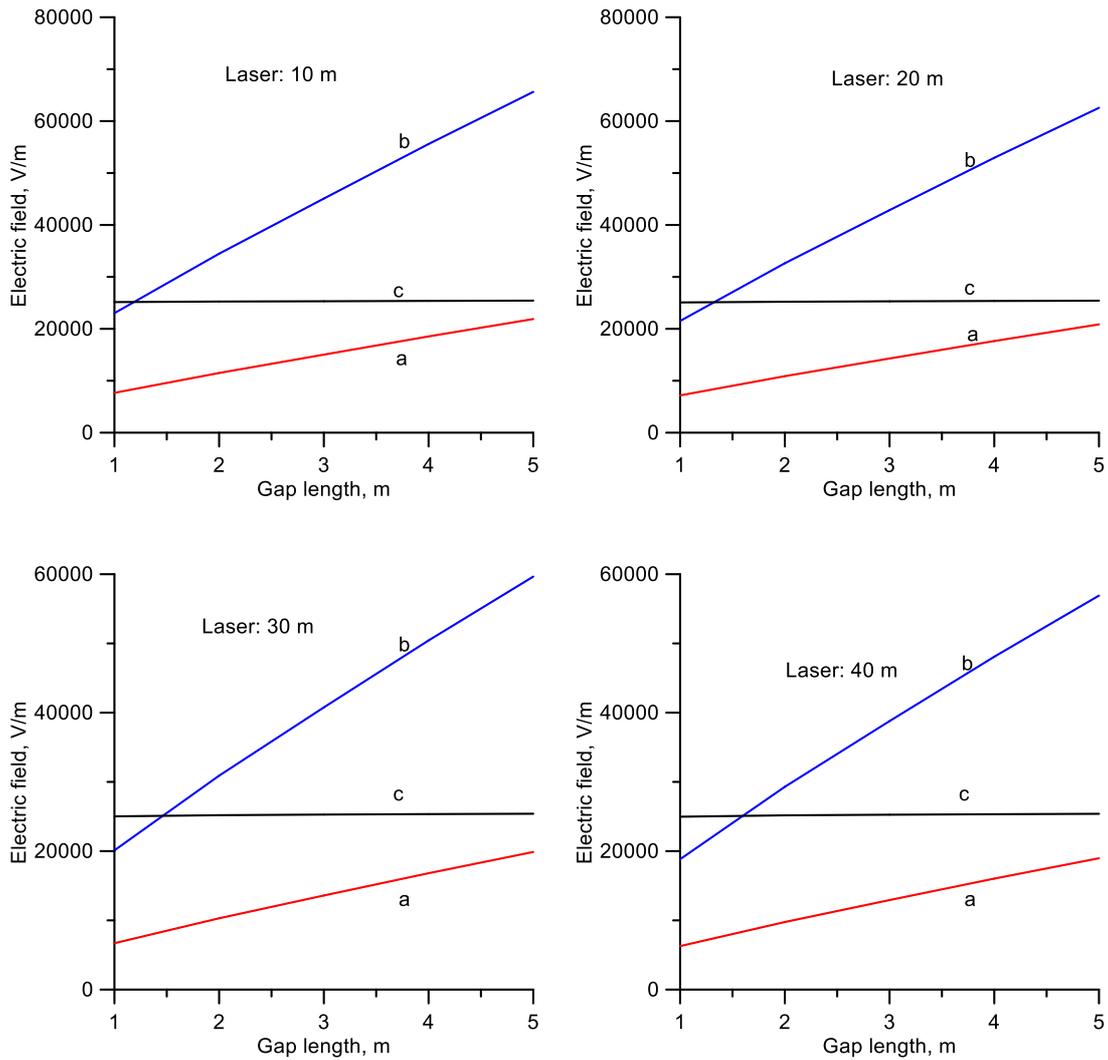

**Extended data Fig. 7. Simulation of the effect of laser filamentation on the lightning flashes initiation**

Magnitude of the background electric fields necessary to initiate laser assisted positive lightning flashes (red curves marked 'a') and laser-assisted negative lightning flashes (blue curves marked 'b'), as a function of the gap length between the lower tip of the laser filament and the tower tip. The curves marked 'c' depict the magnitude of the background electric field necessary for the tower to initiate a negative lightning flash without the assistance of the laser filament. The lengths of the laser filament used in the calculation are given in each diagram. Note that the polarity of the background electric field necessary to initiate a positive lightning flash is opposite to the background electric field necessary to initiate a negative lightning flash. The field presented is not the total field but the background electric field. The absolute electric field at the tower tip is enhanced due to the presence of the tower by a factor of 50 to 100.